\documentclass[aps,pre,twocolumn,superscriptaddress,showpacs]{revtex4}
\usepackage{graphicx}

\begin{document}

\title{Ion beam requirements for fast ignition of inertial fusion targets}

\author{J.J.~Honrubia}
\email{javier.honrubia@upm.es}
\affiliation{ETSI Aeron\'auticos, Universidad Polit\'ecnica de Madrid, Madrid, Spain}

\author{M.~Murakami}
\affiliation{Institute of Laser Engineering, Osaka University, Osaka, Japan}


\date{\today}

\begin{abstract}

Ion beam requirements for fast ignition are investigated by numerical simulation
taking into account new effects such as ion beam divergence not included before.
We assume that ions are generated by the TNSA scheme in a curved foil placed
inside a re-entrant cone and focused on the cone apex or beyond. From the focusing
point to the compressed core ions propagate with a given divergence angle. Ignition
energies are obtained for two compressed fuel configurations heated by proton and
carbon ion beams. The dependence of the ignition energies on the beam divergence
angle and on the position of the ion beam focusing point have been analyzed.
Comparison between TNSA and quasi-monoenergetic ions is also shown.

\end{abstract}


\maketitle

\section{Introduction}

Fast ignition (FI) was proposed 20 years ago as an alternative
to the standard central ignition scheme of inertial fusion
targets \cite{Tabak1994}. It intends to reduce the drive
requirements by separating fuel compression and ignition.
In electron-driven fast ignition (EFI), a fast electron
jet generated by ultra-high intensity lasers triggers
ignition of the thermonuclear fuel. Due to the very
large divergences and the too high kinetic energies
found in EFI experiments and
simulations \cite{Kemp2014,Robinson2014}, ion-driven fast ignition
(IFI) has been taking an increasing interest over the last years.
Ion fast ignition \cite{Tabak1998, Roth2001} offers several
advantages over EFI, such as generation of collimated beams,
well known interaction with the plasma and higher flexibility.
Some examples of such a flexibility are the control of ion spectra
\cite{Fernandez2014,Weng2014}, the possibility of choosing the
optimal ion species \cite{Albright2008,Honrubia2014}, including
deuterium ions \cite{Liu2011}, and the use of multiple beams
for target irradiation \cite{Temporal2006,Temporal2008,Honrubia2014}.
IFI progress so far can be summarized by the experimental
achievement of the required ion energies, high conversion
efficiencies \cite{Brenner2014} and ion beam focusing
\cite{Bartal2012}. This last achievement is crucial because
it points a way forward to get beam focusing into the 30-40 $\mu$m
spots required for IFI. A full review of the current status
of IFI can be found in Ref. \cite{Fernandez2014}.

Just after the first experimental evidence of proton acceleration
by the Target Normal Sheath Acceleration (TNSA) mechanism \cite{Snavely2000},
its application to FI was proposed \cite{Roth2001}. This was followed by
theoretical studies on the TNSA scheme \cite{Mora2003,Murakami2006},
target studies \cite{Hatchett2000,Atzeni2002,Temporal2002,Ramis2004},
new irradiation schemes \cite{Temporal2006,Temporal2008,Honrubia2014} and
the use of ions heavier than protons \cite{Fernandez2009,Honrubia2009}.
Other schemes such as the two-pulse scheme described in Ref. \cite{Tikhon2010}
have been proposed also. This scheme consists of creating a plasma
channel followed by Deuterium-Tritium (DT) acceleration close to
the dense core. Hole boring scalings have been studied recently
in Ref. \cite{Murakami2012}.

Most of the ignition studies carried out so far are based
on rather ideal target configurations and simplified ion
interaction models. For instance, these models do
not take into account cone tip and coronal plasma energy
deposition nor scattering of the beam ions in the hot
plasma between the cone tip and the dense core.

Fully integrated numerical modeling of the IFI scheme, from beam
generation to fuel ignition, is not possible today with
the present computer resources. It requires the
integration of physical processes with very different
spatial and temporal scales. In this paper, we use a
simplified model that assumes an ideal initial distribution
function of ions and computes ion energy deposition
by solving the Fokker-Planck equation. Specifically, we
assume that a highly uniform laser beam, required to
improve the ion beam focusing \cite{Foord2012}, with
irradiances of the order of $10^{20}$ W/cm$^2$ impinges
on a curved foil placed inside a cone, generating a
proton or carbon ion beam at the foil rear surface
by means of the TNSA scheme \cite{Key2006}. Ions are
focused into the inner surface of the cone tip or
beyond by the electrostatic fields generated
near the cone walls. From the beam focusing point,
ions diverge towards the dense fuel. This
paper deals with determining the beam requirements
to ignite a target as a function of the ion
beam divergence taking into account the ion
interactions with the cone tip and their scattering
with the background plasma ions. The goal has
been to obtain realistic estimates of the ignition
energies in the IFI scenario.

This paper is organized as follows. In Section II,
the computational model used in the simulations is
briefly outlined. Details of the multidimensional
Fokker-Planck ion energy deposition model and
its differences with the standard ion tracking
scheme are shown. In Section III, ion energy
deposition and scattering by the cone tip,
energy deposition of divergent beams in the
DT core, ignition energies of proton and carbon
ions with maxwellian and quasi-monoenergetic
energy distributions, and the effect of the position
of the beam focusing point are presented.  Finally,
conclusions and future work are summarized
in Section IV.

\section{Simulation model}

\begin{figure}
\includegraphics[width=.45\textwidth]{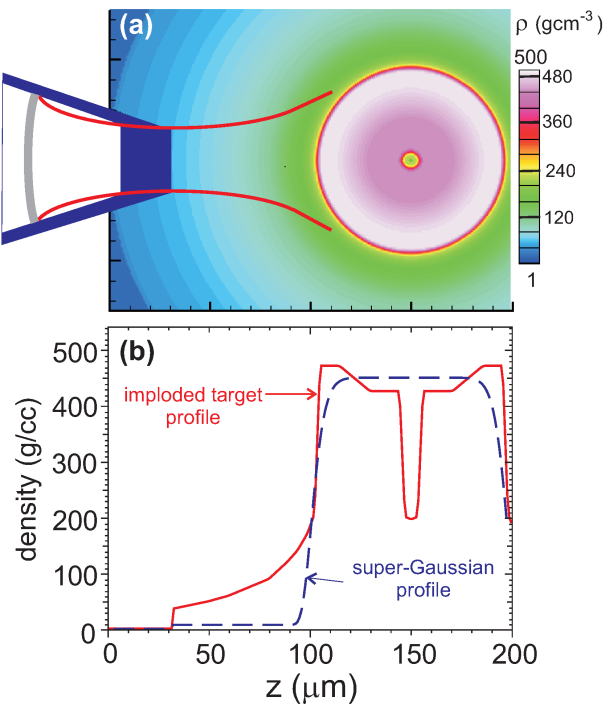}
\caption{\label{fig:1} (a) Sketch of the simulation box used in this work.
Ions are injected at the inner surface of the cone tip sited at z = 5 $\mu$m
and propagate towards the dense core. The distance from the
ion source to the center of the imploded fuel is 650 $\mu$m.
(b) Density profiles along the central line of the simulation box.
The solid line corresponds to the imploded one-shock target of
Ref.~\cite{Clark2007} and the dashed line to an ideal super-gaussian
density distribution.}
\end{figure}

We assume a uniform (constant flux within its cross section)
ion beam impinging on a compressed DT fuel.
The beam is generated by the TNSA scheme \cite{Snavely2000}, which
yields ions with a maxwellian energy distribution. In order to reduce
the ion time spread along the stand-off distance from the generation foil
to the dense fuel, we assume that a re-entrant cone is attached to the
compressed fuel for generating the ion beam as close as possible to the DT.
Two configurations of the compressed fuel at stagnation have been considered.
The first one was taken from the single-shock one-dimensional (1D) target 
design proposed by Clark\&Tabak specifically for fast ignition \cite{Clark2007}.
The simulation box corresponding to this case is shown in Fig.~\ref{fig:1}a.
This fuel configuration is quite ideal because the distortions
induced by the re-entrant cone in the shell during target
implosion and fuel compression are not accounted for. It presents, however,
a large coronal plasma surrounding the core which clearly overestimates
the plasma densities from the cone apex to the DT core when compared with
the density profiles obtained in cone-target simulations \cite{Shay2012}.
As ignition energies depend substantially on the areal density of the coronal
plasma \cite{Honrubia2009}, the 1D imploded target density distribution shown
in Fig.~\ref{fig:1}a allow us to obtain an upper limit of the ignition
energies for the beam conditions studied. The imploded fuel areal density
is $\rho$R = 2.4 gcm$^{-2}$ and we assume an initial temperature of 100 eV
in all the DT with exception of the small central dip, which has a higher
temperature in order to keep the pressure balance with the surrounding
cold and dense fuel. For the initial density and temperature assumed for
the imploded core, 450 gcm$^{-3}$ and 100 eV, respectively, SESAME tables
give a DT pressure of 72 Gbar, which corresponds to a DT adiabat relative
to the Fermi degenerate pressure $\alpha$ = 1.25
($\alpha= p$~(Mbar)$/[2.17~\rho^{5/3}]$ with $\rho$ in gcm$^{-3}$).
A rather low adiabat has been chosen because the target was designed
for a highly isentropic implosion in order to get a quasi-isochoric
stagnated fuel distribution \cite{Clark2007}. Note that the
distance between the inner surface of the cone tip, where the
ion beam is injected, and the core center is 145 $\mu$m and the
distance to the peak density around 100 $\mu$m, which are similar
to those reported by Shay et al. \cite{Shay2012}.

The second configuration of the compressed core is the super-gaussian
density distribution $440\exp[-(r/\delta r)^{12}]$, where $r$ is the
distance to the center, which is placed at $z$ = 150 $\mu$m, and
$\delta r$ = 50 $\mu$m. This distribution is sited on a 10 gcm$^{-3}$
coronal plasma with a negligible areal density, as shown in Fig.~\ref{fig:1}b.
Hence, a lower limit of ignition energies will be obtained in this
case. The initial temperature is again 100 eV. It is worth noting
that the super-gaussian density distribution assumed is similar to
that used in recent electron-driven fast ignition studies
\cite{Honrubia2009b,Johzaki2009,Solodov2009,Strozzi2012}.

Diamond-like-carbon (DLC), aluminum, copper and gold have been chosen as
materials for the cone tip in order to check the importance of ion energy
losses and scattering on the beam energy deposition in the core.
In realistic target designs, the cone should be protected from the x-rays
coming from shell implosion by using heavier materials, such as lead or gold,
at the cone walls. In addition, the cone tip should be thick enough to avoid
rebound shock breakout.

Calculations have been performed with the radiation-hydrodynamics
code SARA, that includes Eulerian hydrodynamics in cylindrical r-z geometry,
flux-limited electron conduction, multigroup radiation transport, ion energy
deposition, DT fusion reactions and $\alpha$-particle transport
\cite{Honrubia1993a, Honrubia1993b}. Ion energy deposition and $\alpha$-particle
transport are computed by the same three-dimensional (3D) Fokker-Planck (FP)
transport module. Equations of state have been taken from SESAME tables
\cite{Sesame}. The numerical parameters used in the simulations have
been chosen as follows: cell widths $\Delta r=\Delta z$ = 1 $\mu$m, initial
time step 3 fs and the number of pseudo-ions injected in the simulation
box was $2\times10^6$.

\subsection{Ion energy deposition}

Ion energy deposition is computed by means of the kinetic Fokker-Planck (FP)
equation. It allows us to include any ion initial distribution function and
the scattering due to ion-ion collisions, neglected in the standard ion
tracking schemes \cite{Temporal2002}. The FP equation is solved in 3D settings
by using a Monte Carlo method similar to that used in our hybrid code for
fast electron transport calculations \cite{Honrubia2006}. Here, for simplicity,
we consider the 1D form of the FP collision term, that for suprathermal
particles reads:

\begin{eqnarray}
\left(\frac{\partial \psi}{\partial t}\right)_{FP}=\frac{\partial(S\psi)}{\partial E}
	+ T \frac{\partial}{\partial \mu}(1-\mu^2)\frac{\partial \psi}{\partial \mu},
\end{eqnarray}
where $\psi=vn$ is the ion velocity $v$ multiplied by the fast ion density
$n(x,E,\mu,t)$ ('particle angular flux' in transport theory), $S$ and $T$ are the
stopping power and mean deflection per unit pathlength, respectively, and $\mu$
is the cosinus of the polar angle of the ion velocity. The collision coefficients
$S$ and $T$ can be written as
\begin{eqnarray}
S(E) &=& \sum_{j} \frac{2\pi q^2 q^2_j m \log \Lambda_j}{m_j E}n_j G(x_j) \\
T(E) &=& \sum_{j} \frac{ \pi q^2 q^2_j   \log \Lambda_j}{2 E^2}n_j H(x_j),
\end{eqnarray}
with
\begin{eqnarray}
  x_j  &=& v / v_j = v / (2\theta_j/m_j)^{1/2}  \\
G(x) &=& erf(x) - \frac{2}{\sqrt{\pi}}x \exp(-x^2)   \\
H(x) &=& \left(1-\frac{1}{2x^2} \right)G(x) + \frac{2}{\sqrt{\pi}}x \exp(-x^2) ,
\end{eqnarray} 
where $j$ stands for plasma species, $E$ is the ion kinetic energy,
$m_j$, $q_j$, $v_j$ and $\theta_j$ are the mass, charge, mean velocity and
temperature of the background plasma species $j$, $m$ and $q$ are the
mass and charge of the fast ion, and $\log \Lambda$ the Coulomb logarithm.

For ion-ion interactions, we have $x_i >> 1$, $G(x_i) \approx$ 1
and $H(x_i)\approx$ 1, while for ion-electron interactions in hot
plasmas such that $v << v_e $, one can take the limit $x_e << 1$. In this limit,
$G(x_e) \approx 4 x_e^3/(3\sqrt \pi)$ and $H(x_e) \approx 4 x_e/(3\sqrt \pi)$.
With the former assumptions, the collision coefficients for an equimolar DT
hot plasma can be written as:
\begin{eqnarray}
  S(E) &=& 2 \pi q^2 e^2 n_{DT} \left[\frac{m}{m_{DT}}\frac{\log\Lambda_i}{E}  + \beta
		  \frac{E^{1/2}}{\theta_e^{3/2}} \log\Lambda_e \right]  \\
  T(E) &=&   \frac{\pi q^2 e^2 }{2} n_{DT} \left[\frac{\log\Lambda_i}{E^2} 
		+ \beta \frac{1}{\theta_e^{1/2}}\frac{1}{E^{3/2}}\log\Lambda_e \right],
\end{eqnarray}
where $\beta=\frac{4}{3\sqrt \pi}\left(\frac{m_e}{m}\right)^{1/2}<<1$, $\theta_e$
and $\theta_i$ are electron and ion temperatures of the background plasma and
$n_{DT}$ the ion density. Note that scattering due to ion-ion collisions is important
for low energies and high electron temperatures $\theta_e$, whereas stopping
with plasma electrons is dominant for high ion energies and low  $\theta_e$.

\begin{figure}
\includegraphics[width=.5\textwidth]{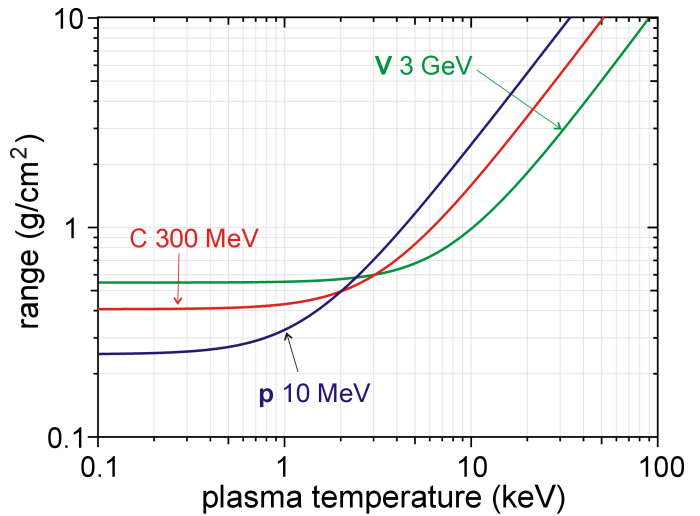}
\caption{\label{fig:2} Range of monoenergetic ions
in DT at 400 gcm$^{-3}$ as a function of the plasma
temperature as obtained from Eq.(2).}
\end{figure}

Ranges of different ion species with typical energies
for IFI are shown in Fig.~\ref{fig:2} as a function of
the DT plasma temperature $\theta$ ($\theta=\theta_e=\theta_i$).
We have used the standard stopping power formula for
classical plasmas, Eq.(2) \cite{Trubnikov, Honrubia1993b}.
This formula predicts range lengthening when plasma electron 
velocities are comparable or higher than the fast ion velocity.
Range lengthening is important for protons with maxwellian energy
distributions generated away from the DT because it balances the
reduction of the ion kinetic energy on target due to time spread, keeping
the ion range almost constant during the ion pulse \cite{Temporal2002}.
It is worth noting that the range lengthening in the temperature
range 0.1 $-$ 10 keV is substantially lower for ions heavier than
protons, as shown in Fig.~\ref{fig:2}. This may be an issue for
quasi-mononenergetic ions, as was pointed out in previous work
\cite{Honrubia2014, Honrubia2009}.

\begin{figure}
\includegraphics[width=.48\textwidth]{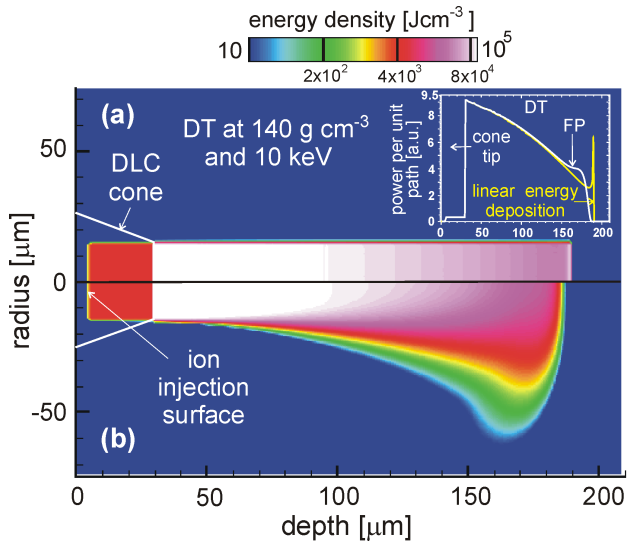}
\caption{\label{fig:3} Comparison of the energy deposition
of a perfectly collimated 10 MeV proton beam in a uniform DT
plasma at 140 gcm$^{-3}$ and 10 keV (a) standard linear energy
deposition model (b) Fokker-Planck (FP) model. Protons are
injected at $z=$ 5 $\mu$m and pass through a 25 $\mu$m
thickness DLC layer at 3 gcm$^{-3}$ mimicking the cone tip.
The energy deposition per unit path length obtaned by both
models are compared in the inset.}
\end{figure}

The energy deposition of a perfectly collimated 10 MeV proton
beam in DT computed by our FP model and by the standard ion
tracking scheme are compared in Fig.~\ref{fig:3}. In both
schemes, the energy deposited per unit pathlength decreases
with the ion penetration because the electron stopping
power is dominant and scales, roughly, as $E^{1/2}$ for
$\theta_i = \theta_e =$ 10 keV. The FP energy deposition
profiles show beam scattering, straggling and blooming
near the end of the range, the absence of the Bragg's
peak and a slight range shortening due to the proton
scattering with the background ions. In the case of ion
tracking, the energy deposition peaks at the end of the
range, showing the well known Bragg's peak, which may
lead to underestimated ignition energies. Note that
the Bragg's peak disappears in the FP energy deposition
profile due to proton scattering. For temperatures lower
than 10 keV, scattering is less important and the
differences between the FP and tracking energy deposition
profiles are lower, but still significant.

To assess the importance
of the FP energy deposition in FI targets, we have compared
the ignition energies of the DT isochoric sphere initially
at 400 gcm$^{-3}$ and 100 eV heated by a uniform cylindrical
beam. Atzeni et al. \cite{Atzeni2002} reported for this target
a minimum ignition energy of 8.5 kJ. Our calculations show the
same ignition energy when the ion scattering is off, while it
increases slightly to 8.75 kJ for the FP energy deposition
model. Thus, despite ion scattering effects seem not to be
important for ion energy deposition, they are not negligible
and should be taken into account in IFI calculations. Heavier
ions, such as carbon, are less sensitive to scattering,
being the ignition energies almost the same with and without
scattering. 

\subsection{Ion pulse on target}

\begin{figure}
\includegraphics[width=.45\textwidth]{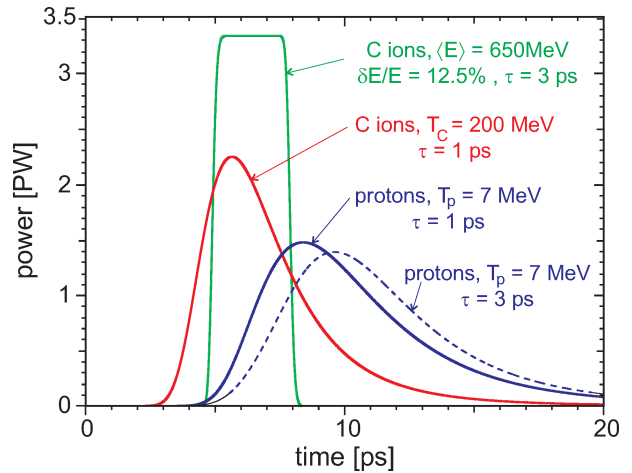}
\caption{\label{fig:4} Beam power at the left surface
of the simulation box as a function of time for different
ion beams with an energy of 10 kJ. The distance from the
ion-generation foil to the left surface of the simulation
box is $d$ = 500 $\mu$m in all cases. $\tau$ stands for
the ion pulse duration on the foil.}
\end{figure}

Beam power on target for proton and carbon ions with maxwellian
energy distributions are plotted in Fig.~\ref{fig:4}. The beam
temperatures are $T_p$ = 7 MeV and $T_C$ = 200 MeV, respectively.
The pulse of quasi-monoenergetic carbon ions with a mean energy
$\langle E \rangle$ = 650 MeV and an energy spread of $\delta E/E$ = 0.125
\cite{Fernandez2014} is also shown for comparison. All beams
have an energy of 10 kJ and are generated at a distance to the
simulation box $d$ = 500 $\mu$m. The power on target $p(t)$ of an
ion beam generated instantaneously in a foil placed at a
distance $d$ is given by the formula \cite{Temporal2002}:
\begin{eqnarray}
p(t)=\frac{8}{3\sqrt{\pi}}\frac{E_b}{t_0}\left(\frac{t_0}{t}\right)^{6}\exp{[-(t_0/t)^2]},
\end{eqnarray}
where $t_0=d(m/2T_b)^{1/2}$, $m$ and $T_b$ are the beam ion mass
and temperature, respectively, $E_b$ the beam energy and $d$ the
distance between the ion-generation foil and the simulation box.
From Eq.(9), the beam power on target can be obtained for a 
pulse of duration $\tau$. Assuming that ions are generated at the
source foil at a constant rate over the pulse duration $\tau$,
the power on target $P(t)$ is given by
\begin{eqnarray}
P(t)=\int_{0}^{\tau} g(t')p(t-t')dt',
\end{eqnarray}
with $t >\tau$ and $g(t)=1/\tau$. This formula has been used
to calculate the pulses shown in Fig.~\ref{fig:4}. We assume
an ion pulse duration of 1 ps throughout this paper, with exception
of the quasi-monoenergetic beams, for which longer pulses, e.g. 3 ps
in Fig.~\ref{fig:4}, are assumed to avoid very high peak powers.
The feasibility of ps pulses for quasi-monoenergetic ion schemes
such as BOA (break-out after burner) or RPA (radiation pressure
acceleration) remains to be demonstrated \cite{Fernandez2014}.

\subsection{Beam focusing}

In addition to the 'conventional' techniques of ion beam
focusing such as ballistic transport \cite{Patel,Key},
magnetic lenses \cite{Schollmeier,Harres,Hofmann}, and
self-generated fields in hollow microcylinders
illuminated by intense sub-picosecond laser pulses
\cite{Toncian}, laser-driven proton beam focusing has been
demonstrated experimentally over the last years. Kar et al.
\cite{Kar} showed beam focusing by using rectangular
or cylindrical hollow lens attached to a foil target.
Offermann et al. \cite{Offermann} found theoretically
and experimentally that ion divergence depends on the
thermal expansion of the co-moving hot electrons, resulting
in a hyperbolic ion beam envelope. Using these results,
Bartal et al. demonstrated experimentally an enhanced
focusing of TNSA protons in cone targets, predicting
spot diameters about 20 $\mu$m for IFI conditions \cite{Bartal2012},
well under the 40 $\mu$m spots required \cite{Honrubia2014}.
The focusing mechanism reported by Bartal et al. is based
on the onset of an electron sheath near the cone walls. This
sheath generates focusing electrostatic fields that avoids
the expansion of co-moving electrons. As these fields are 
generated near the cone walls, they affect mainly to the low
energy ions, keeping the initial divergence of the higher
energy ions generated closer to the beam axis. As a
result, even in the case of a very good focusing
by the cone walls, the beam will follow a hyperbolic
path when traveling towards the core  \cite{Bartal2012,
Foord2012,Qiao2013}. Qiao and coworkers have pointed out
by implicit PIC simulations that the setting up of the
electron sheath at the cone walls may reduce substantially
the laser-to-proton conversion efficiency for the
long pulses required in IFI \cite{Qiao2013}. This can be
mitigated by special designs of cone walls including insulator
materials in order to reduce the electron flow between target
and cone.

Recent experiments carried out at laser intensities of
$\approx 5\times 10^{20}$ Wcm$^{-2}$ relevant for IFI have
evidenced the generation of highly collimated proton beams,
i.e. with divergence full-angles lower than 10$^{\circ}$, in
planar foils illuminated by high-contrast ($\approx 10^{-11}$)
laser beams \cite{Green2014}. This result together with those
reported in Refs. \cite{Bartal2012,Qiao2013}, shows a clear
path to achieve the focusing requirements of IFI if uniform
and high contrast laser beams impinging on a curved target
placed inside a cone are used to ignite the fuel.

As a precise determination of the beam path requires detailed PIC
calculations, we have assumed that ions are focused to the
cone apex or beyond and diverge from there to the compressed core
with a given angle, which is taken as a parameter. 
The reference case calculations shown in the next Section 
assume that the ion beam is focused into a 30 $\mu$m spot
diameter at the inner surface of the cone tip. From this spot,
the ion beam diverges with an opening half-angle $\beta$
as a parameter ranging from 0 to 20$^{\circ}$. For a given $\beta$,
each ion is injected in the simulation box with a randomly selected
initial angle between $0$ and $\beta$. This is an approximation
of the curved ion beam trajectories found theoretically and
experimentally \cite{Offermann}, but allows us to assess the
divergence effects on beam energy requirements. As the
assumption of beam focusing at the inner surface of the cone
tip may appear to be quite restrictive, we have considered
also the cases of ions focused on the outer surface of the
cone tip and on the coronal plasma near the dense DT core.
In these two cases, the divergence effects are much lower
than in the reference case due to the shorter stand-off
distance between the focusing point and the core.

\section{Results}

\subsection{Optimal ion temperature}

\begin{figure}
\includegraphics[width=.45\textwidth]{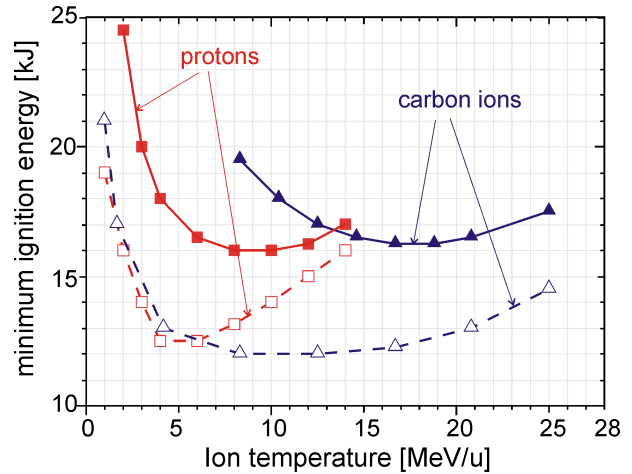}
\caption{\label{fig:5} Ignition energies versus ion temperature
for proton and carbon ion beams impinging perpendicularly on the
target shown in Fig.~\ref{fig:1}. The beams diameter is 30 $\mu$m.
Solid lines correspond to the 1D imploded target density
distribution and dashed lines to the super-gaussian
density distribution.}
\end{figure}

We start by determining the ion temperatures that maximize the beam-target
coupling efficiency or, equivalently, minimize the beam energies required
to ignite the target shown in Fig.~\ref{fig:1}. The results are depicted
in Fig.~\ref{fig:5} for proton and carbon ions with maxwellian energy
distributions. We assume that TNSA carbon ions are generated by using
thin, micrometer scale DLC foils with the rear surface free of protons.
The ignition energies have been obtained as the minimum ion beam energy
for which the thermonuclear fusion power has an exponential or higher
growth sustained in time.

The shape of the curves shown in Fig.~\ref{fig:5} can be explained
as follows. For low temperatures, ions do not penetrate enough in the
target to deposit a significant fraction of their energy into the dense
core, raising the ignition energies. For high temperatures, ions penetrate
into the core more than the optimal areal density of 1.2 gcm$^{-2}$
\cite{Atzeni2002} raising again the ignition energies due to
the increase of the DT mass heated by the beam.

Ignition energies depend on the target density distribution and,
in particular, on the coronal plasma surrounding the core. 
For the 1D imploded density profile with the long coronal plasma shown
in Fig.~\ref{fig:1}b, the ignition energies are minimized for the
'optimal' ion temperatures of $T_p$ = 8 MeV and $T_C$ = 200 MeV
for proton and carbon ion beams, respectively. For the super-gaussian
density profile almost without coronal plasma shown in
Fig.~\ref{fig:1}b, the optimal ion temperatures are $T_p$ = 6 MeV
and $T_C$ = 150 MeV, respectively, substantially lower than those
found for the imploded target density profile. Note also that even
for the optimal ion temperatures, the ignition energies are, roughly,
30\% higher for the imploded target due to the energy deposition
in the coronal plasma. Even in this case, assuming a laser-to-ion
conversion efficiency of 15\% similar to that obtained recently
in proton acceleration experiments \cite{Brenner2014} and close
to the 12\% found by Snavely et al. \cite{Snavely2000}, the laser
beam energy requirements are around 100 kJ. 
 
Independently of the DT density distribution, it is remarkable
that the minimum ignition energies are similar for proton and
heavier ions such as carbon, in agreement with our results for
quasi-monoenergetic beams reported recently \cite{Honrubia2014}.
On the contrary, the optimal ignition energies increase with
the ion atomic number Z scaling, approximately, as $Z^{1.68}$
(obtained for proton, carbon and other ion beams not shown),
which is weaker than the $Z^2$ dependence found for
quasi-monoenergetc ions \cite{Honrubia2014}.
 
In order to simplify the analysis of the next sections, we
assume $T_p$ = 7 MeV and $T_C$ = 200 MeV as 'optimal'
temperatures for the two density profiles depicted in
Fig.~\ref{fig:1}b. Note that these 'optimal' temperatures
are in the plateau region where the ignition energies
depend only weakly on the ion temperature for both density
distributions, as shown in Fig.~\ref{fig:5}.

\subsection{Optimal beam radius}

\begin{figure}
\includegraphics[width=.46\textwidth]{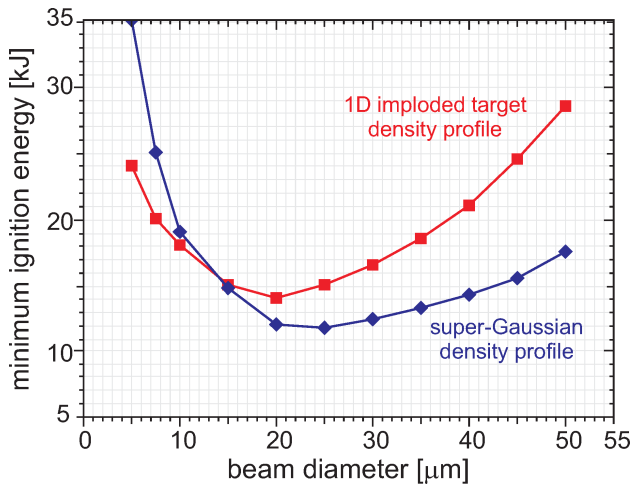}
\caption{\label{fig:6} Ignition energies
of the target shown in Fig.~\ref{fig:1} heated by
a perfectly collimated maxwellian proton beam
with temperature $T_p=$ 7 MeV. Protons are injected
at the inner surface of the DLC cone tip. Red curve
labeled by squares corresponds to the 1D imploded
target density distribution and the blue curve
labeled by diamonds to the super-gaussian density
distribution shown in Fig.~\ref{fig:1}b.}
\end{figure}

In order to determine the ion beam focusing requirements,
we have studied the dependence of the ignition energies
on the beam diameter. A perfectly focused proton beam
with the 'optimal' temperature $T_p$ = 7 MeV impinging
on the target sketched in Fig.~\ref{fig:1} has been chosen
as model problem. The results are shown in Fig.~\ref{fig:6}.
For beams with small diameter ($<$ 10 - 15 $\mu$m,
depending on the density distribution), the energy density
deposited is very high, heating the DT fuel up to
high temperatures with the subsequent range
lengthening of the beam ions. This, together with
the strong hydrodynamic response of the DT plasma to
the high energy deposition may result in ions passing
through the dense core and even escaping by the target
rear surface. As a result, the ignition energies rise
up to high values.

For beams with diameter greater than 35 - 40 $\mu$m,
depending on the density distribution, the ignition energies
rise again due to the increase of the DT volume heated by
the beam. In this case, the ignition energies are sensitive
to the ion energy losses in the coronal plasma, resulting
in a substantial difference between the imploded and
super-gaussian fuel density profiles, as shown in
Fig.~\ref{fig:6}. Thus, for the simulation conditions
analyzed here, beam diameters on DT core should be in
the range between 15 and 35 $\mu$m in order to get
ignition energies close to the minimum value. This range
defines the focusing requirements for IFI and is quite
similar to that obtained in previous works for
quasi-monoenergetic ions \cite{Honrubia2014}. We have
assumed a beam diameter of 30 $\mu$m in the analysis
carried out in the next sections.

\subsection{Energy deposition}

As it was mentioned in the introduction, ions have the advantage
of their classical interaction via Coulomb scattering with the
background plasma. Unlike EFI, collective interactions due
to self-generated fields do not play a significant role in
IFI. This can be explained by the lower current densities
found in IFI, which are about two orders of magnitude lower
than in EFI for the same particle density. In addition,
for ignition-scale targets, temperatures at the cone tip
are high enough to mitigate strongly the growth of resistive
fields. Hence, collective effects such as anomalous ion energy
deposition, beam self-focusing and filamentation of co-moving
electrons may not be important in IFI. Nevertheless, detailed
investigations about those effects in IFI conditions are still
open. A first step in this direction has been given by
Qiao {\it et al.} \cite{Qiao2014}.

\begin{figure}
\includegraphics[width=.46\textwidth]{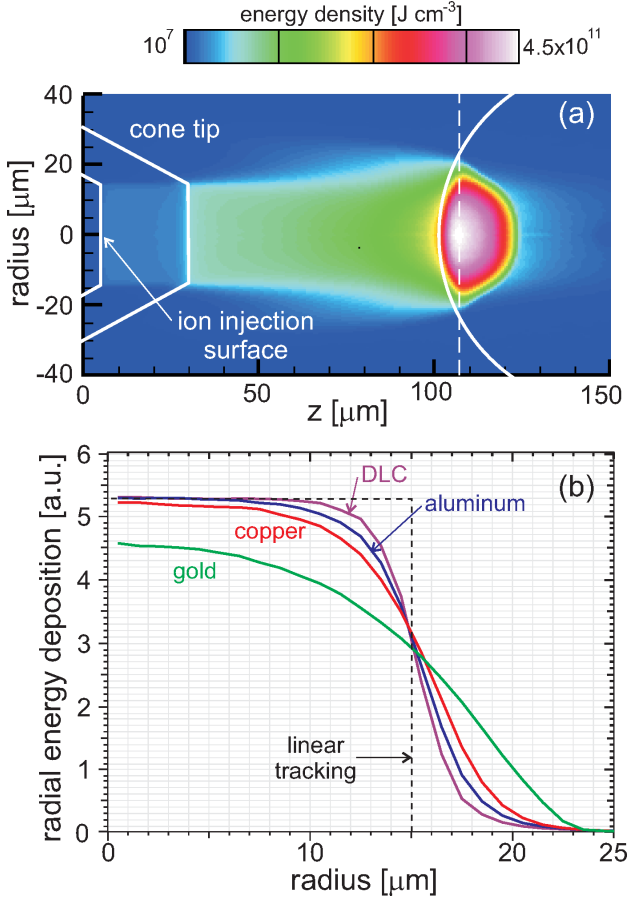}
\caption{\label{fig:7} (a) Energy density isocontours
of a proton beam with a temperature $T_p$ = 7 MeV
impinging perpendicularly on the target shown in
Fig.~\ref{fig:1}a. The solid gold cone tip thickness is
25 $\mu$m. The white circle shows the initial position
of the density isocontour  $\rho=$ 200 gcm$^{-3}$.
(b) Lineouts of the energy density along the dashed
line placed at $z$ = 108 $\mu$m for different cone
tip materials. The lineout corresponding to the
standard proton linear tracking scheme is also shown
for comparison.}
\end{figure}

In addition to the initial ion beam divergence at the
ion-generation foil, ions are scattered when passing
through the cone tip, increasing its divergence \cite{Shmatov2011}.
The effect of the cone tip material on the proton beam energy
deposition is analyzed in Fig.~\ref{fig:7}, where the energy
density deposited by the target of Fig.~\ref{fig:1}a heated
by a proton beam with the 'optimal' temperature $T_p$ = 7 MeV 
is shown. The increase of the beam diameter near the dense
core seen in Fig.~\ref{fig:7}a evidences the proton scattering
by the cone tip. As the mean deflection coefficient $T$ is
proportional to the ion charge squared, Eq.(3),
fast ion scattering is much lower for light materials,
such as DLC or aluminum, as shown in Fig.~\ref{fig:7}b.
On the contrary, for gold cones, scattering is important
resulting in a substantial increase of the beam diameter
near the dense core. For instance, in the case with a
gold cone tip of Fig.~\ref{fig:7}a, the beam radius
increases by around 8 $\mu$m, which is of the same
order than that obtained by Key et al. \cite{Key2006} by
means of the Moliere's multiple scattering theory (which
provides the scattering angle in solid matter, being a
lower limit in plasmas \cite{Shmatov2011}). Moliere's
theory gives a 3.7$^{\circ}$ scattering angle for
monoenergetic protons with energy $(3/2)T_p$ = 10.5 MeV
when traversing a 25 $\mu$m solid gold foil. The
corresponding increase of the beam radius at the
DT core sited at a distance of 75 $\mu$m from the
outer surface of the cone tip is 4.8 $\mu$m, which
is almost a half of the 8 $\mu$m shown in Fig.~\ref{fig:7}b.
This can be explained by the maxwellian, not
mononenergetic, energy distribution of protons and
also by the more effective scattering by plasma ions
than by solid matter. Enhancement of beam radii at
the DT core results in higher ignition energies $E_{ig}$
(see Fig.~\ref{fig:6}). For instance, the ignition
energy of the target sketched in Fig.~\ref{fig:1} is
$E_{ig}$ = 16.5kJ for DLC and aluminum cone tips, whereas
it raises to 17.5 kJ and 21.5 kJ for copper and gold
cones, respectively. Thus, we can conclude that the
use of light materials at the cone tip is more suitable
for IFI, similarly to EFI \cite{Johzaki2009}. A DLC cone
tip is assumed in the reminder of this paper.

\begin{figure}
\includegraphics[width=.49\textwidth]{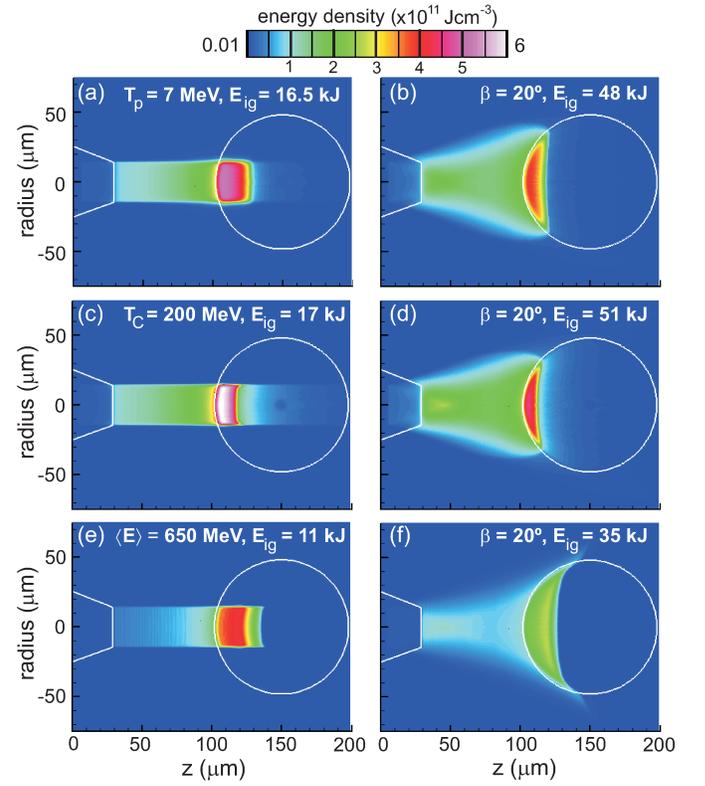}
\caption{\label{fig:8} Maps of the energy density deposited
in the target shown in Fig.~\ref{fig:1}a by (a,b) protons
with $T_p=$ 7 MeV; (c,d) carbon ions with $T_C=$ 200 MeV
and (e,f) 650 MeV carbon ions with an energy spread
$\delta E/E=$ 12.5\%. Plots (a), (c) and (e) correspond
to perfectly collimated ion beams, whereas plots (b), (d)
and (f) correspond to divergent beams with an opening
half-angle $\beta=$ 20$^\circ$. The white curves show
the initial position of the density isocontour
$\rho=$ 200 gcm$^{-3}$. Ignition energies $E_{ig}$ for
each case are shown.}
\end{figure}

The energy deposition by proton and carbon ions with the
optimal temperatures are compared in Fig.~\ref{fig:8} for
perfectly collimated and divergent beams. Energy deposition
by quasi-monoenergetic carbon ions has been included also
for comparison. We analyze first the perfectly collimated
beams, Figs.~\ref{fig:8}a, c and e. Protons with a maxwellian
energy distribution show a more localized energy deposition
than carbon ions with the same distribution. This can be
explained by the better balance between range lengthening
(which is more pronounced for protons than for carbon ions)
and energy spread found for proton beams. This effect is even
more evident by comparing the energy deposition by maxwellian
and quasi-monoenergetic carbon ions, which have a very small
energy spread and, thus, their energy is deposited in a
volume determined mainly by the range lengthening effect.
The higher energy deposition by the maxwellian
ions in the coronal plasma is also remarkable.

Note in Fig.~\ref{fig:8} that the beams have a penetration into
the dense core around 1.5 gcm$^{-2}$, and slightly higher for
the quasi-monoenergetic ions. These values are higher than the
optimal value of 1.2 gcm$^{-2}$
\cite{Atzeni1999,Atzeni2002} due to the energy deposition in the
coronal plasma. The coupling efficiencies, defined as the energy
deposited in the DT at densities higher than 200 gcm$^{-3}$,
for the perfectly collimated beams are 0.58
for the maxwellian proton and carbon beams, and 0.70 for the
quasi-monoenergetic carbon ions, similar to those reported in
Refs.~\cite{Honrubia2009, Honrubia2014}. Coupling efficiencies
rise substantially for the super-gaussian DT density distribution
of Fig.~\ref{fig:1}b due to the lack of the coronal plasma,
reaching the values of 0.86 for the maxwellian proton and carbon
ions, and 0.87 for the quasi-monoenergetic carbon ions. 

The energy deposition maps of beams with a divergence half-angle
$\beta$ = 20$^{\circ}$ are shown in Figs.~\ref{fig:8}b, d and f.
Note the large hot spot elongation due to the high beam divergence assumed,
which distorts substantially the standard rectangular hot spot shape.
As a result, the energy deposition volumes are higher than those found
for collimated beams with the subsequent lower coupling efficiencies
and higher ignition energies. For instance, the coupling efficiencies
are 0.42 for maxwellian proton and carbon ions, increasing up to 0.59
for the quasi-monoenergetic carbon ions. Again, these ions penetrate
deeper in the core, heating a much larger volume than the maxwellian ions.

\subsection{Minimum ignition energies}

\begin{figure}
\includegraphics[width=.46\textwidth]{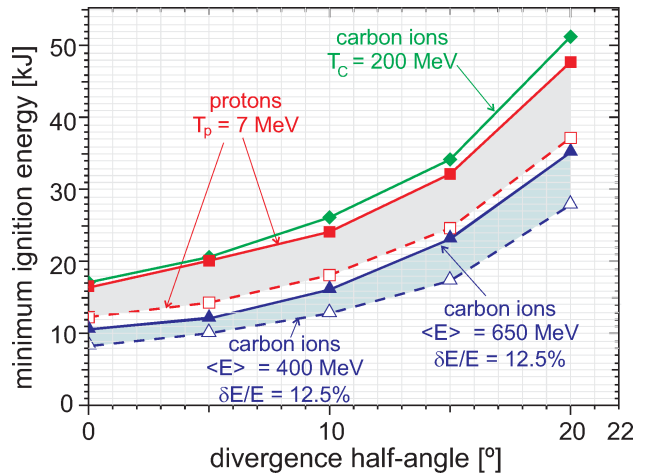}
\caption{\label{fig:9} Ignition energies
of the target shown in Fig.~\ref{fig:1} heated by
maxwellian proton and carbon ions with temperatures
$T_p=$ 7 MeV and $T_C=$ 200 MeV, respectively. The
ignition energies with quasi-monoenergetic carbon
ions are also shown for comparison. Solid lines correspond
to the 1D imploded density profile shown in Fig.~\ref{fig:1}b,
whereas dashed lines correspond to the super-gaussian
density profile. The curves labeled by squares show
the ignition energies for maxwellian protons, the line
labeled by diamonds corresponds to maxwellian carbon
ions, and the lines labeled by triangles correspond to
quasi-monoenergetic carbon ions.}
\end{figure}

The ignition energies for different beams as a function
of the ion divergence half-angle are shown in Fig.~\ref{fig:9}.
These energies have been obtained as described in
Section III.A. As expected, the ignition energies show a
rapid increase with the beam divergence angle. The small
differences between maxwellian proton and carbon ions can
be explained by the lower range lengthening of carbon ions,
which does not balance completely their energy spread along the
source-target distance. It is relevant here to recall that
ignition energies are almost independent of the beam ion atomic
number, provided that ions have the 'optimal' mean kinetic energy,
as can be seen in Fig.~\ref{fig:5}. The much lower ignition
energies found for quasi-monoenergetic ions show the strong
dependence of ignition energies on ion spectrum and the better
coupling of quasi-monoenergetic ions with the plasma.

For maxwellian ions, the solid line curve labeled with solid squares
corresponds to the ignition energies for the 1D imploded target
density profile, while the dotted line curve labeled by empty
squares corresponds to the super-gaussian density profile of
Fig.~\ref{fig:1}b. Both curves show the limit cases of high and
low coronal plasma areal density, respectively. Actually, the
ignition energies of a realistic case would lie in between
these curves, within the grey shaded area of Fig.~\ref{fig:9}.
If we assume a laser-to-ion conversion efficiency of 15\%
\cite{Brenner2014} and an ignitor laser beam energy of 110 kJ,
Fig.~\ref{fig:9} shows that target ignition would be possible
with maxwellian ions if the beam is perfectly collimated or
has a divergence half-angle lower than 8$^{\circ}$ for high
and low coronal plasma areal densities, respectively.

For quasi-monoenergetic ions, the ignition energies are substantially
lower than those obtained for maxwellian ions. They are within the
bottom shaded area of Fig.~\ref{fig:9}. Note that for the case of low
areal density, the kinetic energy of carbon ions has been reduced
from 650 to 400 MeV in order to have the optimal beam-target
coupling.

For the divergence half-angles of 5$^{\circ}$ found in
planar foils experiments \cite{Green2014} and assuming a laser-to-ion
conversion efficiency of 15\%, our results show that it would be
sufficient with an igniting laser beam energy between 130 and 90 kJ
for maxwellian ions, and from 75 to 60 kJ for quasi-monoenergetic
ions, depending on the coronal plasma areal density.

\begin{figure}
\includegraphics[width=.46\textwidth]{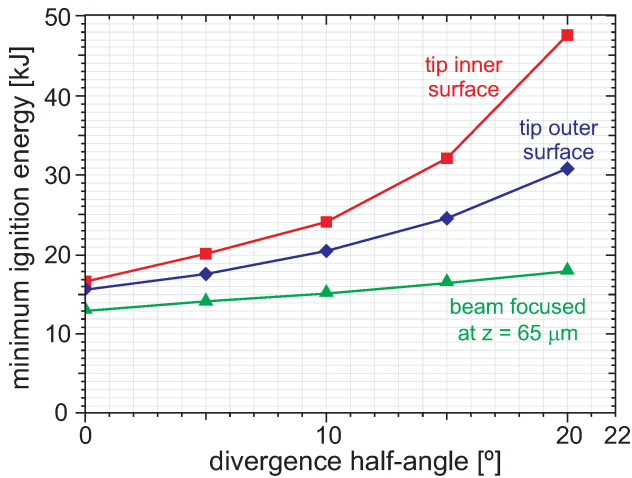}
\caption{\label{fig:10} Ignition energies
of the target shown in Fig.~\ref{fig:1}a heated by
maxwellian protons with temperature $T_p=$ 7 MeV.
Curve labeled by squares shows the ignition
energies for protons injected at the inner
surface of the cone tip. The curve labeled
by diamonds corresponds to protons injected
at the outer surface of the cone tip and the
curve labeled by triangles to protons injected
at $z$ = 65 $\mu$m. In all cases the beam spot
diameter is 30 $\mu$m.}
\end{figure}

\subsection{Beam focusing beyond the cone tip}

In the cases analyzed in Figs. 5 to 9, we have assumed
that the beam is focused by the cone walls on the inner
surface of the cone tip. In this case, the stand-off
distance between the focusing point and the DT
core is maximum and the beam has more room to spread
out. As there is experimental and simulation
evidence of beam focusing beyond the cone tip
\cite{Qiao2013,Qiao2014}, it is worthwhile extending
the study carried out in the last Sections to situations
in which the beam is focused closer to the dense core.
In this case, the DT heating by divergent beams is
more efficient due to the lower beam radius at the core.

In order to analyze the effect of the beam focusing point
on the ignition energies, two additional cases have been
considered: (i) a proton beam focused on the outer surface
of the cone tip and (ii) a proton beam focused 40 $\mu$m away
from the dense core ($z$ = 65 $\mu$m). In both cases, protons
are focused on a 30 $\mu$m diameter spot and have a maxwellian
energy distribution with $T_p$ = 7 MeV. The ignition energies
obtained for the 1D imploded target density profile are plotted
in Fig.~\ref{fig:10}. Note the large reduction of the ignition
energies when the proton beam is focused closer to the DT core.
For instance, if the beam is focused 40 $\mu$m away from the
core, the ignition energies $E_{ig}$ are much lower than in
the reference case and depend weakly on the divergence
half-angle. Even if the beam is focused on the outer
rather than on the inner surface of the cone tip there
is a substantial reduction of $E_{ig}$, specially for
large divergence half-angles.

\section{Conclusions}

Our simulations have evidenced that ion beam ignition energies
have been underestimated so far by using simplified energy 
deposition models and target designs. Fast ion spectrum, focusing
spot diameter and position, beam divergence and ion scattering
with the cone tip are key parameters that determine ignition 
energies. Taking into account all these parameters requires a kinetic
treatment of ion energy deposition such as the linear Fokker-Planck
model used in this work. Thus, our results  generalize those obtained
with the standard linear particle tracking method used so far.

Ignition energies of compressed targets have been evaluated
in the limits of high and low coronal plasma areal densities.
The goal has been to obtain an upper and lower limits of beam
energy requirements for IFI. The results of our study can be
summarized as follows: 

i) The ion beam diameter on the dense core should not exceed
around 40 $\mu$m, which is even higher than the focal spots
inferred from experiments \cite{Bartal2012} and obtained in
large scale PIC simulations of ion acceleration in the IFI
scenario \cite{Qiao2013}. Beam focusing is one of the most
challenging issues of IFI, with high chances of success in
future IFI experiments.

ii) The cone tip material should be as light as possible in order
to reduce ion scattering. Intermediate materials such as copper
are acceptable because they give rise to a modest increase
of the ignition energies when compared with light materials.

iii) As expected, ignition energies depend strongly on the beam
divergence. Our calculations show that for the beam divergence
half-angle of 5$^{\circ}$ measured in the recent experiments,
the ion beam ignition energies of the targets considered here
are from 13 to 20 kJ, depending on the density distribution of
the imploded fuel. These energies are substantially higher than
the standard ignition energies of 10 kJ considered so far.

iv) Assuming the laser-to-proton conversion efficiency
of 15\% found in recent experiments \cite{Brenner2014}, the
laser beam energies required for ignition are in between 90
and 130 kJ for small and large coronal plasma areal densities,
respectively. In principle, these energies could be affordable
with the present laser technology.

v) The position of the ion beam focusing point is crucial to
reduce the ignition energies. If it is beyond the cone tip,
the ignition energies would be reduced substantially and would
be less sensitive to the ion divergence.

vi) If beam focusing were demonstrated for advanced schemes,
such as the break-out afterburner scheme \cite{Fernandez2014},
quasi-monoenergetic ions instead of TNSA ions should be used
due to their better coupling with the compressed fuel.

The challenge over the forthcoming years will be to perform
experiments relevant for a full characterization of ion generation,
focusing and interaction with imploded targets in conditions
close to those found in IFI. The success of IFI will depend on
the possibility of generating ions by means of high contrast,
highly uniform laser beams and focusing the ion beam into
20 - 30 $\mu$m spots as close as possible to the imploded
DT core.

Future studies will include PIC simulations of ion acceleration
and focusing to get a full characterization of the ion source.
This will be followed by integrated simulations similar to
those presented here, but with a self-consistent ion source.

\begin{acknowledgments}
One of the authors (J.J.H.) would like to thank the fruitful
discussions and the hospitality during his stay at ILE.
This work used resources and technical assistance from the
CeSViMa HPC Center of the Polytechnic University of Madrid.
\end{acknowledgments}


\begin{thebibliography}{00}

\bibitem{Tabak1994} M. Tabak. J. Hammer, M.E. Glinsky, W.L. Kruer, S.C. Wilks, J. Woodworth, E.M. Campbell, M.D. Perry and R.J. Mason, Phys. Plasmas {\bf 1}, 1626 (1994).

\bibitem{Kemp2014} A.J. Kemp, F. Fiuza, A. Debayle, T. Johzaki, W.B. Mori, P.K. Patel, Y. Sentoku and L.O. Silva, Nucl. Fusion {\bf 54}, 054002 (2014).

\bibitem{Robinson2014} A.P.L. Robinson, D. Strozzi, J.R. Davies, L. Gremillet, J.J. Honrubia, T. Johzaki, R.J. Kingham, M. Sherlock and A.A. Solodov, Nucl. Fusion {\bf 54}, 054003 (2014).

\bibitem{Tabak1998} M. Tabak and D. Callahan-Miller, Nucl. Instr. Meth. {\bf 415}, 75 (1998).

\bibitem{Roth2001} M. Roth, T.E. Cowan, M.H. Key, S.P. Hatchett, C. Brown, W. Fountain, J. Johnson, D.M. Pennington, R.A. Snavely, S.C. Wilks, K. Yasuike, H. Ruhl, F. Pegoraro, S.V. Bulanov, E.M. Campbell, M.D. Perry and H. Powell, Phys. Rev. Lett. {\bf 86}, 436 (2001).

\bibitem{Fernandez2014} J.C. Fern\'andez, B.J. Albright, F.N. Beg, M.E. Foord, B.M. Hegelich, J.J. Honrubia, M. Roth, R.B. Stephens and L. Yin, Nucl. Fus. {\bf 54}, 054006 (2014).

\bibitem{Weng2014} S.M. Weng, M. Murakami, H. Azechi, J.M. Wang, N. Tasoko, M. Che, Z.M. Sheng, P. Mulser, W. Yu and B.F. Shen, Phys. Plasmas {\bf 21}, 012705 (2014).

\bibitem{Albright2008} B.J. Albright, M.J. Schmitt, J.C. Fern\'andez, G.E. Cragg, I. Tregillis, L. Yin and B.M. Hegelich, J. Phys.: Conf. Ser. {\bf 112}, 022029 (2008).

\bibitem{Honrubia2014} J.J. Honrubia, J.C. Fern\'andez, B.M. Hegelich, M. Murakami and C.D. Enriquez, Laser and Particle Beams {\bf 32}, 419 (2014).

\bibitem{Liu2011} Dong-Xiao Liu, Wei Hong, Lian-Qiang Shan, Shun-Chao Wu and Yu-Qiu Gu, Plasma Phys. Control. Fusion {\bf 53}, 035022 (2011).

\bibitem{Temporal2006} M. Temporal, Phys. Plasmas {\bf 13}, 122704 (2006).

\bibitem{Temporal2008} M. Temporal, J.J. Honrubia and S. Atzeni, Phys. Plasmas {\bf 15}, 052702 (2008).

\bibitem{Brenner2014} C.M. Brenner, A.P.L. Robinson, K. Markey, R.H.H. Scott, R.J. Gray, M. Rosinski, O. Deppert, J. Badziak, D. Batani, J.R. Davies, S.M. Hassan, K.L. Lancaster, K. Li, I.O. Musgrave, P.A. Norreys, J. Pasley, M. Roth, H.-P. Schlenvoigt, C. Spindloe, M. Tatarakis, T. Winstone, J. Wolowski, D. Wyatt, P. McKenna and D. Neely, App. Phys. Lett. {\bf 104}, 081123 (2014).

\bibitem{Bartal2012} T. Bartal, M.E. Foord, C.Bellei, M.H. Key, K.A. Flippo, S.A. Gaillard, D.T. Offermann, P.K. Patel, L.C. Jarrott, D.P. Higginson, M. Roth, A. Otten, D. Kraus, R.B. Stephens, H.S. McLean, E.M. Giraldez, M.S. Wei, D.C. Gautier and F.N. Beg, Nat. Phys. {\bf 8}, 139 (2012).

\bibitem{Snavely2000} R.A. Snavely, M.H. Key, S.P. Hatchett, T.E. Cowan, M. Roth, T.W. Phillips, M.A. Stoyer, E.A. Henry, T.C. Sangster, M.S. Singh, S.C. Wilks, A. MacKinnon, A. Offenberger, D.M. Pennington, K. Yasuike, A.B. Langdon, B.F. Lasinski, J. Johnson, M.D. Perry, and E.M. Campbell, Phys. Rev. Lett. {\bf 85}, 2495 (2000).

\bibitem{Mora2003} P. Mora, Phys. Rev. Lett. {\bf 90}, 185002 (2003)

\bibitem{Murakami2006} M. Murakami and M.M. Basko, Phys. Plasmas {\bf 13}, 012105 (2006).

\bibitem{Hatchett2000} S.P. Hatchett et al., {\it Cone-focused fast ignition: Sub-Ignition Proof-of-Principle Experiments}, (Ed. M. Key) contribution to the 6th Workshop on Fast Ignition of Fusion Targets , 16-18 November2002, St. Petes Beach, Florida, USA (2002).

\bibitem{Atzeni2002} S. Atzeni, M. Temporal and J.J. Honrubia, Nucl. Fus. {\bf 42}, L1 (2002).

\bibitem{Temporal2002} M. Temporal, J.J. Honrubia and S. Atzeni, Phys. Plasmas {\bf 9}, 3098  (2002).

\bibitem{Ramis2004} R. Ramis and J. Ram\'irez, Indirectly driven target design for fast igniton with proton beams, Nucl. Fusion {\bf 44}, 720 (2004).

\bibitem{Fernandez2009} J.C. Fern\'andez, J.J. Honrubia, B.J. Albright, K.A. Flippo, D.C. Gautier, B.M. Hegelich, M.J. Schmitt, M. Temporal and L. Yin, Nucl. Fus. {\bf 49}, 065004 (2009).

\bibitem{Honrubia2009} J.J. Honrubia, J.C. Fern\'andez, M. Temporal, B.M. Hegelich and J. Meyer-ter-Vehn, Phys. Plasmas {\bf 16}, 102701 (2009).

\bibitem{Tikhon2010} V.T. Tikhonchuk, T. Schlegel, C. Regan, M. Temporal, J.-L. Feugeas, Ph. Nicola\"i and X. Ribeyre, Nucl. Fusion {\bf 50}, 045003 (2010).

\bibitem{Murakami2012} S.M. Weng, M. Murakami, P. Mulser and Z.M. Sheng, New J. Phys. {\bf 14}, 063026 (2012).

\bibitem{Foord2012} M.E. Foord, T. Bartal, C. Bellei, M. Key, K. Flippo, R.B. Stephens, P. K. Patel, H. S. McLean, L.C. Jarrott, M.S. Wei and F.N. Beg‭, Phys. Plasmas {\bf 19}, 056702 (2012).

\bibitem{Key2006} M.H. Key, R.R. Freeman, S.P. Hatchett, A.J. MacKinnon, P.K. Patel, R.A. Snavely and R.B. Stephens, Fusion Sci. and Tecnol. {\bf 49}, 440 (2006).

\bibitem{Clark2007} C.D. Clark and M. Tabak, Nucl. Fus. {\bf 47}, 1147 (2007).

\bibitem{Shay2012} H.D. Shay, P. Amendt, D. Clark, D. Ho, M. Key, J. Koning, M. Marinak, D. Strozzi and M. Tabak, Phys. Plasmas {\bf 19}, 092706 (2012).

\bibitem{Honrubia2009b} J.J. Honrubia and J. Meyer-ter-Vehn, Plasma Phys. Control. Fusion {\bf 51}, 014008 (2009).

\bibitem{Johzaki2009} T. Johzaki, Y. Nakao and K. Mima, Phys. Plasmas {\bf 16}, 062706 (2009).

\bibitem{Solodov2009} A.A. Solodov, K.S. Anderson, R. Betti, V. Gotcheva, J. Myatt, J.A. Deletrez, S. Skupsky, W. Theobald and J.C.Stoeckl, Phys. Plasmas {\bf 16}, 056309 (2009).

\bibitem{Strozzi2012} D.J. Strozzi, M. Tabak, D.J. Larson, L. Divol, A.J. Kemp, C. Bellei, M.M. Marinak and M.H. Key, Phys. Plasmas {\bf 19}, 072711 (2012).

\bibitem{Honrubia1993a} J.J. Honrubia, J. Quant. Spectrosc. Radiat. Transfer {\bf 49}, 491 (1993).

\bibitem{Honrubia1993b} J.J. Honrubia, {\it Charged particle transport} in Fusion by Inertial Confinement, G. Velarde, Y. Ronen and J.M. Mart\'inez-Val (Eds.), CRC Press, Boca Rat\'on, Florida (1993).

\bibitem{Sesame} SESAME: The Los Alamos National Laboratory Equation of State Database, LA-UR-92-3407 (1992).

\bibitem{Honrubia2006} J.J. Honrubia, C. Alfons\'in, L. Alonso, B. P\'erez and J.A. Cerrada, Laser Part. Beams {\bf 24}, 217 (2006).

\bibitem{Trubnikov} B.A. Trubnikov, {\it Particle interactions in a fully ionized plasma} in Review of Plasma Physics, M.A. Leontovich (Ed.), Vol. 1, Consultants Bureau, New York (1965).

\bibitem{Patel} P.K. Patel, A.J. Mackinnon, M.H. Key, T.E. Cowan, M.E. Foord, M. Allen, D.F. Price, H. Ruhl, P.T. Springer and R. Stephens, Phys. Rev. Lett. {\bf 91}, 125004 (2008).

\bibitem{Key} M.H. Key, Phys. Plasmas {\bf 14}, 055502 (2007).

\bibitem{Schollmeier} M. Schollmeier, S. Becker, M. Gei$\beta$el, K.A. Flippo, A. Blazevic, S.A. Gaillard, D.C. Gautier, F. Gruner, K. Harres, M. Kimmel, F. Nurnberg, P. Rambo, U. Schramm, J. Schreiber, J. Schutrumpf, J. Schwarz, N.A. Tahir, B. Atherton, D. Habs, B.M. Hegelich, and M. Roth, Phys. Rev. Lett. {\bf 101}, 055004 (2008). 

\bibitem{Harres} K. Harres, I. Alber, A. Tauschwitz, V. Bagnoud, H. Daido, M. Günther, F. Nürnberg, A. Otten, M. Schollmeier, J. Sch\"utrumpf, M. Tampo and M. Roth, Phys. Plasmas {\bf 17}, 023107 (2010).

\bibitem{Hofmann} I. Hofmann, J. Meyer-ter-Vehn, X. Yan, A. Orzhekhovskaya and S. Yaramyshev, Phys. Rev. ST Accel. Beams {\bf 14}, 031304 (2011).

\bibitem{Toncian} T. Toncian, M. Borghesi, J. Fuchs, E. d'Humi\`eres, P. Antici, P. Audebert, E. Brambrink, C.A. Cecchetti, A. Pipahl, L. Romagnani and O. Willi, Science {\bf 312}, 410 (2006).

\bibitem{Kar} S. Kar, K. Markey, P.T. Simpson, C. Bellei, J.S. Green, S.R. Nagel, S. Kneip, D.C. Carroll, B. Dromey, L. Willingale, E.L. Clark, P. McKenna, Z. Najmudin, K. Krushelnick, P. Norreys, R.J. Clarke, D. Neely, M. Borghesi, and M. Zepf, Phys. Rev. Lett. {\bf 100}, 105004 (2008).

\bibitem{Offermann} D.T. Offermann, K.A. Flippo, J. Cobble, M.J. Schmitt, S.A. Gaillard, T. Bartal, D.V. Rose, D.R. Welch, M. Geissel and M. Schollmeier, Phys. Plasmas {\bf 18}, 056713 (2011).

\bibitem{Qiao2013} B. Qiao, M.E. Foord, M.S. Wei, R.B. Stephens, M.H. Key, H. McLean, P.K. Patel and F.N. Beg, Phys. Rev E {\bf 87}, 013108 (2013).

\bibitem{Green2014} J.S. Green, N.P. Dover, M. Borghesi, C.M. Brenner, F.H. Cameron, D.C. Carroll, P.S. Foster, P. Gallegos, G. Gregori and P. McKenna, Plasma Phys. Control. Fus. {\bf 56}, 084001 (2014). 

\bibitem{Qiao2014}  B. Qiao et al., Acceleration, {\it Focusing and Energy Deposition of High Intensity Proton Beams in the kilojoule multi-picosecond Laser Regime for Fast Ignition}, presented at the 2$^{nd}$ International Conference on High Energy Density Physics, Beijing, China, September 21 - 24 (2014).
 
\bibitem{Shmatov2011} M.L. Shmatov, Laser Part. Beams {\bf 29}, pp. 339-344 (2011).

\bibitem{Atzeni1999} S. Atzeni, Phys. Plasmas {\bf 6}, 3316 (1999).


\end{thebibliography}
\end{document}